\documentclass[compsoc,conference,a4paper,10pt,times]{IEEEtran}
\pdfoutput=1

\usepackage{cite}
\usepackage{amsmath,amssymb,amsfonts}
\usepackage{algorithmic}
\usepackage{graphicx}
\usepackage{textcomp}
\usepackage{xcolor}
\usepackage{listings}
\usepackage{booktabs}
\usepackage{subcaption}
\usepackage{float}

\def\BibTeX{{\rm B\kern-.05em{\sc i\kern-.025em b}\kern-.08em
    T\kern-.1667em\lower.7ex\hbox{E}\kern-.125emX}}

\definecolor{backcolour}{rgb}{0.95,0.95,0.92}
\definecolor{codegreen}{rgb}{0,0.6,0}
\definecolor{codegray}{rgb}{0.5,0.5,0.5}
\definecolor{codepurple}{rgb}{0.58,0,0.82}

\lstdefinestyle{codeStyle}{
    backgroundcolor=\color{backcolour},   
    commentstyle=\color{codepurple},
    keywordstyle=\color{blue},
    numberstyle=\tiny\color{codegray},
    stringstyle=\color{codegreen},
    basicstyle=\ttfamily\footnotesize,
    breakatwhitespace=false,         
    breaklines=true,                 
    keepspaces=true,                 
    showspaces=false,                
    showstringspaces=false,
    showtabs=false,                  
    tabsize=2
}    

\newcommand{\code}[3]{
    \lstinputlisting[
        label={lst:#1},
        language=#2,
        style=codeStyle,
        caption=#1,
        captionpos=b
    ]{#3}
}    

\usepackage[colorlinks=true,urlcolor=black]{hyperref}
\def\BibTeX{{\rm B\kern-.05em{\sc i\kern-.025em b}\kern-.08em
    T\kern-.1667em\lower.7ex\hbox{E}\kern-.125emX}}
\begin{document}

\title{Threadbox: Sandboxing for Modular Security}

\author{\IEEEauthorblockN{Maysara Alhindi}
\IEEEauthorblockA{\textit{University of Bristol}\\
Bristol, UK\\
maysara.alhindi@bristol.ac.uk}
\and
\IEEEauthorblockN{Joseph Hallett}
\IEEEauthorblockA{ \textit{University of Bristol}\\
Bristol, UK\\
joseph.hallett@bristol.ac.uk}
}

\maketitle

\begin{abstract}
There are many sandboxing mechanisms provided by operating systems to limit what resources applications can access, however, sometimes the use of these mechanisms requires developers to refactor their code to fit the sandboxing model. In this work, we investigate what makes existing sandboxing mechanisms challenging to apply to certain types of applications, and propose Threadbox, a sandboxing mechanism that enables having modular and independent sandboxes, and can be applied to threads and sandbox specific functions. We present case studies to illustrate the applicability of the idea and discuss its limitations.
\end{abstract}


\section{Introduction}

You can not achieve security with complexity. A sandboxing solution that requires developers to redesign their code or write very long security policies, is likely not to be adopted. Different operating systems provide sandboxing mechanisms that allow developers to restrict the resources an application can access. Linux has Seccomp which operates on the system call interface and enables filtering what system calls an application can make. OpenBSD has Pledge which is similar to Seccomp, but it provides an abstraction layer that makes it easier and more simple to use, while FreeBSD has Capsicum that restricts access to file descriptors and the global namespace.

In their work, Alhindi et al. examined the adoption of these mechanisms in software packages and noticed that sometimes these mechanisms can be truly challenging to apply and require developers to refactor their code~\cite{alhindi2024sandboxing}.

For example, Firefox implemented sandboxing later than most mainstream web browsers. The problem with sandboxing Firefox was in its monolithic process model where different components of the browser are run under the same process. This design made it hard to implement the principle of the least privilege and confine the process to only the needed resources using the current sandboxing solutions~\cite{Maass-thesis}. Solving this problem required decoupling the browser components such as the renderer and the plugin system to impose a distinct sandbox on addons while allowing for a different sandbox for the rest of the browser. Without such separation, both the browser and addons would share the same sandbox, offering limited protection against addons compromising the browser~\cite{Maass-thesis}.

The current version of Firefox uses Seccomp, Pledge, and Capsicum to implement sandboxing, the case can be made that Firefox is a very complex software, and as a consequence, it was challenging to sandbox it, however, other types of software are less complex than Firefox but still require refactoring work to accommodate sandboxing mechanisms~\cite{alhindi2024sandboxing}, for instance, the developer of the \texttt{gmid} package had to refactor to it to a multi-process model and implement inter-process communication to be able to sandbox the software using the different sandboxing mechanisms~\cite{Omar_Polo_2021}. What makes these mechanisms require refactoring to be used in many types of applications? And how well do they fit the requirements of common software such as web servers and GUI applications?

We investigate the challenges of current sandboxing mechanisms and propose \emph{Threadbox} as a model of sandboxing that combines some of Pledge's and Seccomp's features into a solution that provides \emph{modular} sandboxing that is easy to apply to programs without the need to re-architecture them. Similar to Seccomp, Threadbox applies the sandbox to threads but does not inherit the sandbox to children threads, and has a syntax similar to Pledge's. Threadbox can also be used to sandbox Python functions by annotating functions with the needed permissions, making it easy to sandbox smaller components of applications.

\code{Sandboxing functions using Threadbox}{Python}{code/threadboxexample.py}

We make the following research contributions:
\begin{itemize}
\item We investigate the existing sandboxing mechanisms (Seccomp, Pledge, and Capsicum) and explore what makes them challenging to apply to common software such as web servers. We demonstrate that current solutions sometimes require developers to refactor their code, and we investigate the technical reasons behind this.
\item We propose Threadbox, a sandboxing mechanism that allows for having \emph{modular} sandboxing that can be attached to threads. We explain its design and how it compares to current solutions, and discuss the model of sandboxing functions and its~benefits.
\item To demonstrate the applicability of Threadbox, we present 3 case studies where Threadbox is used to sandbox software without the need for refactoring.
\end{itemize}

This work identifies challenges in current models and proposes a model that provides a layer of security focusing on modularity and usability.

\section{Motivation and Existing Solutions}

\subsection{What does it take to sandbox a web server?}

\begin{figure*}[hbt]
    \centering
    \begin{subfigure}[t]{0.3\textwidth}
        \centering
        \includegraphics[height=1.2in]{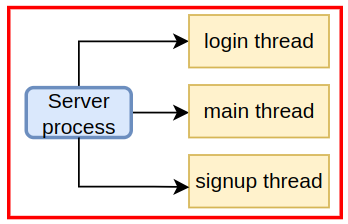}
        \caption{Attaching a sandbox to the whole process}
        \label{fig:server_sandbox_main}
    \end{subfigure}%
    ~
    \begin{subfigure}[t]{0.3\textwidth}
        \centering
        \includegraphics[height=1.2in]{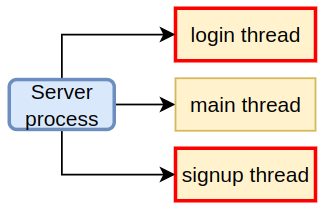}
        \caption{Sandboxing threads using Seccomp}
        \label{fig:server_sandbox_seccomp} 
    \end{subfigure}%
    ~
    \begin{subfigure}[t]{0.3\textwidth}
        \centering
        \includegraphics[height=1.2in]{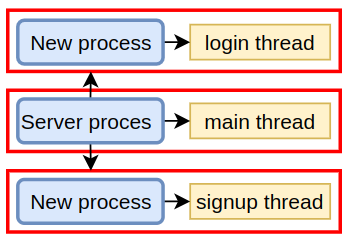}
        \caption{Using a multi-process model and attach a sandbox to each process}
        \label{fig:server_sandbox_processes}
    \end{subfigure}%
    ~
    \caption{Sandboxing Flask web server using different sandboxing models}
    \label{fig:diff_sandboxes}
\end{figure*}

Web applications are considered a prime target for sandboxing since they are exposed to a wide range of attacks, handle users' data, easily accessible and widely spread. Many web development frameworks enable developers to utilize thousands of packages that make developing web applications easier. The availability of these packages creates an environment where developers can quickly implement web applications in a relatively short period, however, this reliance poses a real security threat as many incidents involved attackers developing packages that contain malicious code. There have been documented cases of harmful \texttt{npm} and \texttt{pip} modules that engage in actions like extracting user data from forms, executing shell injection attacks, and installing malware and crypto miners \cite{duan2020towards, vaidya2019security}. The increased harm associated with these attacks stems from the fact that web applications and their dependencies have unrestricted access to a wide range of resources, including the file system, the internet, and environment variables. Sandboxing can help in reducing the attack surface of web applications and mitigating the risks posed by malicious packages, but how difficult it is to apply Pledge, Seccomp, or Capsicum to such~applications?

Let us consider sandboxing a Flask web application. Sandboxing a Flask web server is an interesting case, as many development frameworks such as Flask, Django, and Spring create new threads for each request the server receives and close these threads once the request handler finishes execution~\cite{flask-doc}. To sandbox such a model using Capsicum, Pledge, or Seccomp, many approaches can be taken as shown in Figure \ref{fig:diff_sandboxes}. The easiest of which would be to allow the cloning and networking system calls, and file-system access for the main process of the application, since each request would require launching a new thread, communicating over a socket, and reading template files. This also means that every request handler will have access to these permissions even if they are not needed in the handler's code, as requests are launched in threads in the same process and will be under the process sandbox. Combining these permissions is dangerous, especially when applied to all parts of the application. This allows malicious packages and attackers to launch a new process, and communicate over the network. However, this model does not require the developer to do any re-architecturing of their codebase and can be fairly easy to implement, for instance, doing such a sandbox with Pledge can be done in one line of code.

Another approach would be to create and launch a new process for each request handler. This would allow for fine-grained sandboxing where developers can create specific sandboxes for each process (every request handler will be launched in a new process). In Pledge's case, this would also mean that the main process can be sandboxed since the Pledge model does not inherent sandboxes to child processes, and independent sandboxes can be applied to each process. However, this is not as easy as it seems due to many reasons. First, having a new process for each request is not practical as it will lead to huge memory footprints, second, launching new processes for each request is not advised in many frameworks such as Flask since these frameworks are designed to do quick CRUD operations, and launching a new process for each request would increase the response time.

Unlike Pledge and Capsicum, Seccomp filters are applied to threads, which makes it easier to sandbox the handlers' code without the need for moving into a multi-process model. However, because Seccomp filters are inherited by child threads and processes, this means that the main thread, the one that spawns request handlers' threads, will be left without a sandbox, because if sandboxed, the exact filters will be applied to every child thread, and it would not be possible to create specific sandboxes for each request handler. Also, this means that if a request handler launches an additional thread or a process, the Seccomp filters of the request handler will be applied to these sub-threads and processes which might cause problems since these sub-threads might need more permissions than their parent. It is common for web developers to launch sub-processes to leverage the many programs operating systems provide~\cite{Abbadini_Facchinetti_Oldani_Rossi_Paraboschi_2023}. This practice makes using Seccomp to create a sandbox for requests handlers challenging as the sandbox have to both account for the resources of the handler and the sub processes. Furthermore, a simple Python program that only prints a string will trigger hundreds of system calls. This makes sandboxing a Flask server using Seccomp challenging since developers have to manually type and restrict many system calls, which could lead to errors and can be hard to maintain.

So what do we take from this? Having a sandbox that applies to the whole process can be suitable for many types of applications where the process resources can be accessed before applying the sandbox and where the process is mainly doing one type of task. For instance, sandboxing \texttt{tcpdump} using Capsicum is a straightforward task since the needed files and resources are opened and initialized, and after that, the process goes into the capability mode and performs the main loop under the sandbox restrictions~\cite{capsicum-cap-for-unix}. However, in many cases such as Flask web applications, the per-process model does not work well and re-architecturing would be needed to have granular sandboxing. The web application case is not the only case where the multi-threading paradigm does not fit the per-process sandboxing model. Another case is GUI applications, GUI applications use multithreading to be able to visually present interfaces, simultaneously handle user input and perform logic in the background. Many frameworks for GUI applications are designed this way and refactoring these applications to be multi-process to fit sandboxing is not easy nor practical.

Seccomp model applies to threads which makes it more flexible to apply to multi-threaded applications, however, the inheritance part does not allow the creation of specific sandboxes for all parts of the application, and re-architecturing might be needed to protect all components. In addition, Seccomp syntax requires the developers to manually type long lists of system calls and consider different architectures and libraries versions when doing so, a task that is very demanding and requires an understanding of system internals.

\subsection{Precision Protection}

Pledge model of sandboxing offers a simpler approach for developers to adopt sandboxing when compared to Seccomp, but both Pledge and Seccomp models have their challenges. Sandboxing mechanisms that can not be easily applied to specific components lead to situations like in Firefox, where a substantial amount of work had to be done to accommodate sandboxing or lead to the sandboxing mechanism not being adopted enough by developers. In their research, Masses et al. commented on the Firefox sandboxing problem by suggesting that effective sandboxing mechanisms should be designed to be easily applied to a specific problem (Add-ons in the case of Firefox) and should be simple enough to use without having neither over nor under-constrained sandboxes~\cite{Maass-thesis}.

In the light of this, we designed Threadbox, a sandboxing mechanism that is designed to be applied with modularity in mind. Threadbox has a hybrid model that merges some of Seccomp and Pledge features into a solution that can be easily applied to software like Flask. Threadbox attaches sandboxes to threads, but it does not propagate the sandbox to children processes or threads, and has a simpler syntax similar to Pledge's. With Threadbox, sandboxing all the components of Flask web server becomes possible and as easy as writing one line of code to sandbox each handler, and the main thread.

\begin{figure}[H]
    \centering
    \includegraphics[scale=.42]{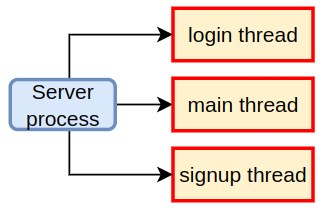}
    \caption{Sandboxing request handlers and the main thread using Threadbox}
    \label{fig:thread_sandbox}
\end{figure}

\subsection{Function-based sandboxing}

Modularity promotes security. In general, the more modular a program is, the easier it becomes to manage and maintain it. Modularity also makes sandboxing easier and more effective, since a separate sandbox can be applied to each component confining it to what it only needs. What if we could sandbox specific functions? Figuring out what permissions a whole process needs can be a challenging task, especially if the process is complex, and with mechanisms like Seccomp, this becomes even harder since every system call the process makes has to be considered. Functions have specific input and output, and perform specific task, which makes it easier and less intimidating to know what permissions they need. What makes functions a perfect target to apply a sandbox is that most applications already are split into different functions, and it becomes a much easier task to have modular sandboxing and to consider a sandbox for a specific function rather than the whole application, specially that it is a good practice to write functions to do one specific task and to break down complex logic into smaller functions. 

Alhindi et al. conducted a usability study on Seccomp, and noticed that some participants tried to implement the sandboxing policy for specific functions and had a function-driven mindset \cite{alhindi2025playing}. Threadbox builds on that concept as it allows sandboxing specific functions with just one line of code.

\section{Threadbox Architecture}

\subsection{Design Overview}

Threadbox is a sandboxing mechanism to sandbox specific threads in a process. It has two components, a kernel component which is an LSM (Linux Security Module) that assigns permissions to threads and restricts what a sandboxed thread can do. It also has a user-space component that defines APIs for different programming languages to be able to communicate the sandbox configurations to the kernel component. We also developed features to help developers inspect the sandbox while it is running and debug it when the sandbox is violated.

Threadbox has a hybrid model that blends some of Seccomp and Pledge features into one solution. Table~\ref{tab:threadbox-compare} shows how Threadbox compares with other mechanisms. Similar to Seccomp, Threadbox applies to threads, however, the sandbox filters are not propagated to sub-threads or processes. Threadbox adopts a policy similar to promises in Pledge, where it hides the complexity of the system calls interface behind an abstraction layer that is easy to understand and examine. While in reality, there is no \emph{the best} sandboxing mechanism, as all of them offer distinct features and are designed for different purposes, our aim was to have a mechanism that is easy to use, does not require refactoring, can be applied to specific modules, and fits many common applications that are difficult to sandbox using existing mechanisms.

\begin{table}
    \centering
    \caption{Comparison between Threadbox and other sandboxing mechanisms}
    \begin{tabular}{lllll}
      \toprule
      Feature & Seccomp & Pledge & Capsicum & Threadbox \\
      \midrule
      Model & Thread & Process & Process & Thread \\
      Inheritance & Yes & No \footnotemark & Yes & No \\
      Policy abstraction & No & Yes & No & Yes \\
      \bottomrule
    \end{tabular}
    \label{tab:threadbox-compare}
\end{table}

\subsection{User-space component}

When designing Threadbox, we decided to adopt a model similar to Pledge model where there is a layer of abstraction. With Seccomp, developers have to manually type each system call they want to allow or deny, and have to inspect the calls their program makes using external tools like strace, contrary to the Pledge model, where developers can easily determine the needed permissions based on the program functionality, if it uses networking, the \texttt{inet} promise will be required and no need to check every system call the program makes. The list of system calls can also be different across various libc implementations which will require developers extra work to make their sandbox portable. Not all developers know the concept of system calls and would be challenging for them to determine the needed permissions for their program without a level of abstraction like in Pledge where system calls and permissions are grouped under simple terms that can be easily understood. Threadbox is intended to be used in web frameworks, GUI applications and other scenarios where developers might not necessarily be familiar with system internals. \footnotetext[1]{Unless \texttt{execpromises} parameter is set.}

Similar limitations apply to the Capsicum model where developers still need to know and inspect how their program functions under the hood to be able to sandbox it, for instance, developers can restrict IOCTL and FCNTL requests their program issues, but simple Python programs issue a lot of such requests under the hood. Similarly, using Capsicum to restrict access to file descriptors can be challenging if the programming language does not expose the internals of file access to the developer. Another limitation is that the Capsicum model is intended to be used where the resources of the program are opened, and then the program will enter the capability mode where it would not be able to access the global namespace anymore. This model fits many cases but in some instances, it would require developers to do a lot of refactoring to be able to sandbox their applications.

Despite the abstraction layer the Pledge model has, many of its promises still need expertise to be understood. For instance, if developers want to use Pledge to sandbox a GUI application, they have to use the \texttt{unix} promise, and while for some it might be obvious why GUI applications need to use \texttt{unix} sockets to function, many developers might not know this. Similar limitations have been highlighted before in AppArmor policies where users found it difficult to understand many of its permissions despite it having an abstraction layer ~\cite{schreuders2012towards}. We introduced a few changes to Pledge promises to make them simpler and a better fit for our model of sandboxing. Threadbox has only 7 promises (Pledge has 33). Table \ref{tab:threadbox-promises} shows how Threadbox promises compare to Pledge's.

We removed a lot of promises that as shown in Alhindi et al. work were rarely used. This would further simplify the sandboxing experience since developers do not need to consider reading or testing if their program uses these promises. We also removed the \texttt{stdio} promise and allowed its permissions by default, since most the packages examined in Alhindi et al. adoption study that used Pledge allow the \texttt{stdio} promise. We combined the \texttt{cpath} and \texttt{dpath} promises into the \texttt{wpath} promise, as both of these promises result in modifying the file system by either creating files or nodes, and thus can be simplified and included under the \texttt{wpath} promise that allows modifying the file system. The \texttt{unix} promise is renamed to either \texttt{gui} or \texttt{ipc} promise, as this would make it more understandable to a wider audience of developers. We also added the \texttt{threading} promise that allows programs to only launch new threads without being able to fork new processes, some ports of the Pledge system call added this promise as well \cite{Tunney_pledge_2022, serenityos_github}.

In Pledge, promises are not just a direct mapping to a group of system calls, they are more than that. For instance, the \texttt{dns} promise allows operating on sockets but also it allows reading the \texttt{/etc/resolv.conf} file without explicitly allowing the \texttt{rpath} promise, since this file is needed in DNS queries. In Threadbox, this is not the case. Threadbox promises are just mappings to LSM hooks that are related to the promise category. Implementing a more detailed model like in Pledge is difficult in Linux, as the kernel space and user space are developed by different teams, where there is a wide variety of user-space programs implementations. For instance, certain implementations of libc might use \texttt{/etc/resolv.conf} file when doing a DNS query, other implementations might use something else.

\begin{table}
    \centering
    \caption{Comparison between the promises of Threadbox and those of OpenBSD Pledge.  Threadbox simplifies Pledge's promises, merging several related promises into a single Threadbox~promise.}
    \label{tab:my-table}
    \begin{tabular}{ll}
        \toprule
        Threadbox & Pledge \\
        \midrule
        \texttt{proc} & \texttt{proc exec} \\ 
        \texttt{rpath} & \texttt{rpath} \\ 
        \texttt{wpath} & \texttt{wpath dpath cpath tmppath} \\
        \texttt{net} & \texttt{inet dns} \\ 
        \texttt{id} & \texttt{id} \\ 
        \texttt{gui/ipc} & \texttt{unix} \\
        \texttt{threading} & -  \\
      \bottomrule
    \end{tabular}
    \label{tab:threadbox-promises}
\end{table}

We developed libraries that provide APIs to use Threadbox. These libraries provide an interface to place threads in a sandbox and have three functions as shown in Listing~\ref{lst:Threadbox functions}. The \texttt{sandbox\_ps} function takes no arguments and does not impose any restrictions on the calling process, it only declares that this process will sandbox one of its threads at some point. While this function is not \emph{necessary}, it improves the performance of the sandbox as the logic in the kernel code is only checked for processes that are in a sandbox and not every process in the system. The \texttt{permissions} function puts the calling thread into a sandbox and takes a string argument that represents the promises a thread intends to use, and two boolean arguments, debug and complain, we will expand on these arguments later in the debugging section. The \texttt{sandbox\_function} takes the same arguments as the \texttt{permissions} function and places the annotated function in a sandbox (Only for Python functions).

These arguments are communicated to the LSM via files in the security virtual file system. Once the system boots, the LSM creates four files, \texttt{sandbox\_ps}, \texttt{promises}, \texttt{debug}, and \texttt{complain}. Any user program that wishes to use Threadbox can simply write to these files. The libraries we developed are no more than a wrapper that communicates the sandbox settings to the kernel via these files.

\code{Threadbox functions}{Python}{code/libfunctions.py}

When placing threads into a sandbox, we make sure that the sandboxed code can not gain privileges more than its parent by setting \texttt{PR\_SET\_NO\_NEW\_PRIVS} property to the sandboxed thread. Denying system calls can lead to security flaws such as privilege escalations, the famous \texttt{sendmail} bug is an example of why setting the \texttt{PR\_SET\_NO\_NEW\_PRIVS} property is important~\cite{Jake_2010}. While in our model, the process will be killed if a system call is denied, and thus such attacks will not occur, we thought that setting this property can provide an extra layer of protection without affecting the performance or functionality.

\section{Kernel-space component}

\subsection*{Why an LSM?}

Linux Security Modules enable developers to place hooks in the kernel to restrict what programs can do. Hooks can be placed on actions that affect or access kernel objects, and many of the system calls have corresponding LSM hooks. A hook that relates to a system call will be triggered whenever that system call is executed. Developers can inspect the arguments of the system call and either allow its execution to continue or stop it. The hooks can also access data about the context in which a system call is run and can inspect data such as the calling process and~thread.

One aspect that makes developing LSMs challenging is that they have to be compiled into the kernel and can not be plugged in without rebuilding the kernel and rebooting the system, which is time-consuming and makes debugging harder. Unlike LSMs, Linux Kernel Modules (LKM) can be inserted at run-time. Hooking system calls can be done with LKMs using \texttt{ftrace}, a technique that is created mainly for debugging purposes, however, using \texttt{ftrace} to hook system calls is not optimal and results in significant performance issues when compared to LSMs, therefore, we decided to create an LSM to build the kernel component as LSMs provide a standard interface to the kernel and are designed to impose security policies. Many security mechanisms are developed as LSMs, such as SELinux, AppArmor, Smack, Landlock, and many more.

There are projects that port the Pledge system call to Linux systems, such as the work by Tunney that used Seccomp to implement Pledge \cite{Tunney_pledge_2022}, and the work by Kling \cite{serenityos_github} that implemented the Pledge system call in SerenityOS. Using Seccomp (Like Tunney's implementation) in our case to implement the sandbox enforcement is not possible because Seccomp inherits the sandbox filters to sub-threads and processes and can not achieve the goals of our design. Kling's implementation of Pledge is process-based and is almost identical to the original implementation in OpenBSD. Threadbox is different from these implementations as it is an LSM, thread-based with no inheritance, simplifies Pledge promises, offers a unique debugging and automation interface, and proposes the model of sandboxing specific-functions.

\subsection*{Threadbox LSM implementation}

When writing kernel code, one must be very careful, a simple bug in the kernel could lead to damaging hardware, affect all users, or faulting the whole system, especially since kernel modules and LSMs are written in C and simple mistakes could lead to severe consequences and thus, any buffer or memory addresses received from the user-space should be carefully checked. Developers should be mindful when using resources like memory in the kernel space, a poorly performant kernel component could slow the whole system, therefore, cleaning up and removing any unused resources becomes critical. That is why developers are advised to keep their modules as simple as possible, and to implement features in the user space when applicable, focusing the kernel code on doing only necessary and specific tasks~\cite{corbet2005linux}.

Threadbox LSM is only about 600 lines of code, 200 of which are the main logic of the sandbox, and the rest is related to placing hooks and initialization code. The code of Threadbox is open source and still in early development~stages\footnote{https://github.com/null9900/threadbox}.

In the kernel space, processes and threads are represented by the same struct, the \texttt{task} struct. This struct contains many fields including the \texttt{tid} and \texttt{tgid} where \texttt{tid} is the thread ID and \texttt{tgid} is the ID of the thread group which maps to the concept of a process. In the kernel space, processes are a group of threads and a process with one thread will have the same \texttt{tid} and \texttt{tgid}, and additional threads will share the same \texttt{tgid} and have a different \texttt{tid}. Creating processes and threads trigger the same LSM hook which is \texttt{task\_alloc}, and when a process or a thread dies, the \texttt{task\_free} hook is triggered. 

We map threads to promises in the kernel in a simple struct that has few fields such as the \texttt{tid}, \texttt{tgid}, and an integer field to keep track of what promises each thread is allowed to use. The struct is shown in Listing~\ref{lst:Thread struct}. We store both the \texttt{tid} and \texttt{tgid} of the sandboxed thread because thread IDs are not guaranteed to be unique across different processes, so to be able to correctly identify a sandboxed
thread by its \texttt{tid}, we have to also know its \texttt{tgid} (process ID). The struct also contains fields that store whether the thread is in debugging or complain mode and the debugging name associated with it. Each sandboxed thread gets a struct assigned to it, and the struct gets cleared when the thread exits.

\code{Thread struct}{C}{code/thread_struct.c}

We utilized a similar approach to OpenBSD's implementation of the Pledge system call where the declared promises are stored as a number, with each promise having a unique number (ID) that is checked by bit-shifting the promise ID and doing a bit-wise OR on the declared promises by the thread. The code that checks if the required promise is granted for a thread is implemented as a C Macro that is called within every LSM hook with the needed promise. Certain hooks require different promises based on their parameters, for instance, the \texttt{socket\_bind} hook requires the \texttt{net} promise if the socket is of an INET type, while the \texttt{gui} promise will be needed if the socket in question is a UNIX socket. Listing \ref{lst:The task alloc LSM hook} shows the \texttt{task\_alloc} hook in Threadbox, the clone flags parameter is examined to determine if the new task is a thread or a process, and then the related required promise is checked.

Our LSM defines many hooks to check whether to allow or disallow system calls, we also hook \texttt{task\_free} event which is triggered when a process or a thread dies, we do this to make sure to free up the resources and remove any related data when a sandboxed thread exits.

Once a hook is triggered, we check if the calling process is using Threadbox, and then we check if the calling thread is sandboxed, and if it is, then we check if it has sufficient promises for the LSM hook to run. Our design does not impose any restrictions on threads in a process by default unless the thread is put in a sandbox since LSM hooks are checked only for threads that are sandboxed. Figure \ref{fig:flow_diagram} in the Appendix shows a simplified flow diagram of the Threadbox LSM, focusing on what happens when the permissions function is called and when a system call is triggered.

\code{The task alloc LSM hook}{C}{code/lsm_hook.c}

Threadbox LSM utilizes the virtual file system in Linux to be able to communicate with the user-space programs. Once the operating system boots, the LSM creates four virtual files, \texttt{sandbox\_ps}, \texttt{promises}, \texttt{debug}, and \texttt{complain}. A user-space program that writes anything to \texttt{sandbox\_ps} file simply tells the kernel that this process is going to use Threadbox for one of its threads. To sandbox a thread, a program can write the list of promises to the
\texttt{promises} file, and the LSM will check the calling thread and assigns the communicated promises to it in the kernel. After that, the system calls that are performed by the calling thread will be checked in the LSM against the previously declared promises. Threads are allowed to declare promises only once. Any attempt to use a permission outside the promised ones will cause the process to be killed and exit, thus preventing attackers from resuming or repeating their attacks again. If a
thread is in a sandbox, any subsequent writes to \texttt{promises} will not have any effect. 

\section{Sandboxing Python Functions}

If we want to look at the most basic building blocks of an application, functions, is it possible to enforce a sandbox to a specific function? Programs are usually split into functions, each function performing a well-defined task, thus, placing a sandbox on functions can be easier than sandboxing the whole program, because functions have a well-defined input and output and no need to refactor the program to make it modular, since programs already consist of different modules, functions.

The concept of a function is only visible to the language interpreter or the compiler, and it is difficult to distinguish a function in the kernel space. In the kernel space, threads are considered to be the smallest unit of processing, thus, if we want to have kernel-enforced sandboxing, threads would be the smallest unit that is possible to sandbox. Unlike processes, threads share the same memory space and are not expensive to create and destroy. To be able to apply kernel-enforced sandboxing to specific functions, one would need to launch these functions in separate threads.

Python provides the \texttt{functools} module, which allows for creating function annotations, where developers can write code to be run before and after annotated functions. We used this module to create an annotation \texttt{sandbox\_function} that would launch the annotated function into a new thread and apply the Threadbox sandbox for it. This makes sandboxing specific functions as easy as annotating the function with the needed permissions. The way that the \texttt{sandbox\_function} call
works is that it launches the annotated Python function into a new thread, places it into a sandbox with the promises provided as an argument to the \texttt{sandbox\_function} call, and then waits for the launched thread to finish execution and returns the results. The code behind \texttt{sandbox\_function} annotation is shown in Listing ~\ref{lst:Sandbox function annotation wrapper}.

\code{Sandbox function annotation wrapper}{Python}{code/sandbox_function.py}

\section{Debugging and Helper Tools}

The design of LSMs does not provide an easy way to establish a direct feedback loop with user-space programs. For instance, when a system call fails, LSMs return an \texttt{-EPERM} error code, but this error code is very generic and can result from other system components such as violated capabilities. Also, a simple error code does not enable developers to provide enough information to allow debugging, such as what system call has been denied.  While developing the LSM, we noticed that different programs respond differently to this error code, and some even output error messages that can be very hard to trace back to the LSM rule being violated. 

Providing debugging information via user-space files is not a practical solution. Reading or writing to user-space files from the kernel space is very uncommon and highly discouraged. As discussed earlier, a kernel component should be kept very simple, handling files usually involves parsing and interpreting data from the user space, and this process can introduce many bugs. Furthermore, figuring out where a file resides in user space is not that simple, since files can exist in different namespaces. The most efficient way to deliver feedback when the sandbox is violated is via kernel logs. LSMs can write to \texttt{/var/log/dmesg} using functions like \texttt{printk}, and users can scan such logs either by using the \texttt{dmesg} utility or by reading the kernel log file. Many LSMs utilize this approach such as~AppArmor.

With Threadbox, developers can supply a name to sandboxed threads using the \texttt{debug} parameter to the \texttt{permissions} or \texttt{sandbox\_function} functions. Whenever a Threadbox sandbox is violated, the \texttt{dmesg} log will output the violated promise and associate it with the named thread in the logs as shown in Listing~\ref{lst:Threadbox kernel log}, which makes it easy to search for the thread and determine what permissions have been denied. This is particularly useful when sandboxing multi-threaded applications, as many sandboxed components can be running in parallel, and it would very challenging to determine what component has its sandbox violated without being able to attach identifiable information with each sandbox (thread).

We also provided a complain mode similar to the one in AppArmor. When setting the complain argument to True in the \texttt{permissions} or \texttt{sandbox\_function} functions, the sandboxed thread will not be killed when the sandbox is violated, and the LSM will log every promise that has been used by that thread along with the thread name until it finishes~execution.

\code{Threadbox kernel log}{Lisp}{code/kernel_log.txt}

Threadbox outputs error messages that specify the violated permissions and can be easily traced back to the bugged components. The complain mode can also help developers in figuring out what permissions (promises) each function or thread needs.

\section{Case studies}

The code of these case studies, alongside demonstration videos are available on GitHub. \footnote{https://github.com/null9900/threadbox-evaluation}

\subsection{Case Study: Sandboxing Web Development Frameworks}
\label{sec:sec01}

This case study focuses on an open-source Flask web application (flask-datta-able) that implements a user dashboard. The app connects to a database and provides authentication functionality, allowing users to register, log in, and log out. It also serves various static files based on the requested page \cite{flask_dashboard}.

As noted earlier, it is common for developers to rely on third-party packages to implement functionality. In the Python ecosystem, the \texttt{pip} repository hosts thousands of open-source libraries. This case study highlights one such library: PyYAML, a widely used YAML parser downloaded millions of times per day \cite{pyyaml}. Notably, PyYAML has a known vulnerability—CVE-2017-18342—that allows command injection if untrusted input is passed to the \texttt{load} function. To mitigate this, the maintainers introduced a safer alternative, \texttt{safe\_load}. However, the vulnerable \texttt{load} function remains available, and many developers remain unaware of the distinction, as some widely viewed StackOverflow posts still recommend using \texttt{yaml.load} over \texttt{yaml.safe\_load}~\cite{pyyaml_question}.

To simulate a real-world vulnerability, we modified the Flask application to allow users to upload YAML files during the registration process. We then used the vulnerable \texttt{yaml.load} function to parse these files. This setup mirrors a realistic threat scenario involving a widely used application pattern and a common form of vulnerable code, exactly the kind of problem Threadbox is designed to address. Flask applications typically include multiple endpoints that handle different kinds of data, each of which may benefit from a distinct sandbox policy. This case study demonstrates how Threadbox enables modular sandboxing while also effectively containing the vulnerable library to prevent system compromise.

In this case study, our goal is to sandbox the different request handlers in a Flask web application. So, how does Threadbox help us achieve that?

The Flask app has three different request handlers to manage login and registration, \texttt{/login}, \texttt{/register}, and \texttt{/logout}. Listing \ref{lst:Flask login handler} shows the login request handler. This handler accepts POST and GET requests on the \texttt{/login} endpoint. It fetches the user name and the password and checks for the user in the database, verifies its password, and then sends a response back to the user. When the login page is requested by a user, a new thread will be launched, this function will be executed, and then the thread will be destroyed. 

With Threadbox, developers can assign specific permissions to each handler. To determine the necessary permissions for a given handler, developers can assign a unique name to the function and run Threadbox in learning mode by setting the third argument in the \texttt{permissions} function to True. In this mode, Threadbox observes the handler’s behaviour and logs which permissions it requires. Listing~\ref{lst:Learning mode output for the login endpoint} shows the output generated when the login handler is triggered under learning mode. As seen in the logs, the login handler only needs two permissions: \texttt{rpath} (for reading files) and \texttt{net} (for network access).

\code{Flask login handler}{Python}{code/flask_login.py}

\code{Learning mode output for the login endpoint}{Lisp}{code/flask_learning_logs.txt}

By adding one line of code \texttt{permissions("net rpath")} at the top of the function as shown in Listing~\ref{lst:Flask login handler}, the login handler is sandboxed. The login handler only needs to be able to have read-only access to files (reading HTML templates in this case) and to be able to access the networking socket to send responses and access the database, thus the \texttt{net} and \texttt{rpath} promises should be allowed. If the handler accesses any promises outside the defined ones, the web server process will be killed by the kernel and the incident will be reported in the log files. Developers can assign permissions to each request handler by calling the \texttt{permissions} function with needed~promises.

Backend endpoints often process different kinds of data, and therefore require different permissions. For example, the \texttt{/logout} endpoint in this case study only requires network access, while the \texttt{/login} and \texttt{/register} handlers need networking and filesystem read access to serve static content.
With process-based sandboxing, either all endpoints are granted the union of all required permissions, resulting in over privileged code, or developers must refactor the application to spawn a new process for every request, each with its own specific sandbox. Listing \ref{lst:Flask logout handler} shows the \texttt{/logout} endpoint. Threadbox allows for attaching a different sandbox for each endpoint without refactoring.

\code{Flask logout handler}{Python}{code/flask_logout.py}

With this model, since each sandbox is independently applied to threads without inheritance, developers can also sandbox the main thread, in this case, \texttt{threading unix net rpath} promises are needed to sandbox the main thread. The \texttt{net} promise is to allow the app to operate on sockets, the \texttt{unix} promise is needed to communicate with the \texttt{systemd-resolved} service to resolve DNS queries, the \texttt{rpath} promise is required to read template and styles files, and the \texttt{threading} promise because the main thread launches new threads for each request handler.

While the sandbox for the main thread is relaxed, it is better than no sandboxing at all. The main thread is still not able to open files with write-access, can not launch new processes, and can not manipulate users on the system. Our approach does not promise full security, for instance, if the login handler thread is compromised, the attacker will still be able to issue read and networking system calls, but will not be able to launch new processes/threads, write files, operate on local sockets, or manipulate users' data. This limits the damage an attacker can do in case of a compromise, especially in the case of a malicious library.

The registration endpoint handler receives user data, including the username, email, password, and a YAML file. It parses this data, checks if the user already exists, and if not, creates a new user. The registration handler is shown in Listing~\ref{lst:Flask register handler}.

Similar to the login handler, the registration handler only needs two permissions: \texttt{rpath} and \texttt{net}. However, this handler uses the vulnerable method \texttt{yaml.load}. To test if Threadbox can prevent exploitation, we tested the registration endpoint by uploading two files: one valid YAML file and one malicious file containing a command injection exploit. The malicious file contains the command shown in Listing \ref{lst:Contents of a malicious file that would result in command injection in PyYAML}, which would disclose the contents of the \texttt{/etc/passwd} file.

\code{Flask register handler}{Python}{code/flask_register.py}

\code{Contents of a malicious file that would result in command injection in PyYAML}{C}{code/yaml_exploit.txt}

When uploading the valid file, the Threadbox sandbox does not interfere, as parsing this file does not require more permissions than those granted to the handler. However, when uploading the malicious file, Threadbox kills the entire server process, as it detects that the registration thread attempts to use the \texttt{proc} promise to execute the injected command, which is not allowed. The logs when uploading the malicious file are shown below:

\code{Threadbox output when triggering the PyYAML exploit in the register handler}{Lisp}{code/yaml_exp_output.txt}

\subsection{Case Study: Sandboxing Python Functions}

As a case study, we sandboxed \texttt{magic-wormhole}, an open-source Python package that provides both a CLI and a library to send files, directories, and text from one computer to another securely~\cite{wormhole_github}. The project has over 21,000 stars on GitHub. This case study focuses on sandboxing the CLI component of the package and aims to demonstrate how Threadbox can be used to sandbox individual functions.

To test the security guarantees offered by Threadbox, we implemented a simulated malicious component (backdoor) that launches a new process and executes commands, but only if the file being transferred has a specific name.

This case study highlights a unique capability of Threadbox: the ability to sandbox individual functions. This function-level sandboxing model enables complete modularity by allowing developers to attach independent sandboxes to specific functions. Each sandbox can be finely tailored to restrict the function to only the permissions it actually requires, thereby minimizing the attack surface and containing the impact of any malicious or vulnerable code.

Let us set a goal here: we want to sandbox the CLI component of \texttt{magic-wormhole}, focusing on the parts responsible for receiving data. One particular function, \texttt{\_extract\_file}, is in charge of extracting received directories, and is shown in Listing~\ref{lst:The function responsible for extracting compressed files}. When a user sends a directory, \texttt{magic-wormhole} compresses it, and extracts it on the other end using \texttt{\_extract\_file} function. This function is a good candidate for sandboxing, as vulnerabilities have frequently arisen from extracting malicious compressed files. Many decompression libraries have had vulnerabilities before, including the very recent \texttt{xz} backdoor~\cite{lins2024critical}.
So, how do we sandbox this single function using Threadbox? And how does Threadbox simplify the sandboxing experience for developers?

\code{The function responsible for extracting compressed files}{Python}{code/wormhole_dir_extract.py}

Threadbox allows developers to attach sandboxes to specific functions. To do this, developers can annotate the function with the \texttt{@sandbox\_function} decorator and specify the required permissions. To determine those permissions, developers can assign a name to the sandbox and enable learning mode as follows: \texttt{@sandbox\_function(" ", "Extract file", True)}. When running the function under learning mode, the logs reveal that the only required promise is \texttt{wpath}. As shown in Listing~\ref{lst:The function responsible for extracting compressed files}, sandboxing this function is as simple as adding one line of code.

Because the function is small and self-contained, developers can also approach this by simply reading its code and inferring the required permissions. In this case, extracting a directory and writing the output to the file system should only require the \texttt{wpath} promise.

If developers want to sandbox other functions, they can annotate each one individually with Threadbox, enable learning mode, and assign a unique name for each sandbox. After executing the functions, Threadbox will report the required promises in the logs and associate them with the named sandboxes. In comparison, determining the necessary system calls for multiple components with Seccomp would be extremely challenging and time consuming.

Mechanisms like Pledge or Capsicum limit developers to create sandbox policies for the entire application, which means surveying the entire codebase and testing multiple execution paths. Not as simple as considering one function. Threadbox’s function-level model empowers developers to sandbox one function at a time and evolve their security posture incrementally, making it easier to sandbox applications.

When using Threadbox, the best approach would be to target functions that perform specific tasks, where it is easy to figure out the needed permissions. For instance, the function \texttt{\_parse\_offer} in \texttt{magic-wormhole} is responsible for figuring out whether the received data is text, file, or folder, and then it calls the needed functions to parse the received data as shown in Listing ~\ref{lst:Parse offer function in magic-wormhole}. Sandboxing this function requires the \texttt{wpath} promise since downloading files and folders would write to the file system, however, if the communicated data was text, this would not be needed. Sandboxing this function would result in giving \texttt{wpath} permission to a code that does not need it.

\code{Parse offer function in magic-wormhole}{Python}{code/wormhole_parse_offer.py}

A better approach would be to sandbox the \texttt{\_handle\_file}, \texttt{\_handle\_directory}, \texttt{\_write\_file}, \texttt{\_write\_directory}, and \texttt{\_handle\_text} functions separately rather than attaching a sandbox to their parent function. The \texttt{\_handle\_file}, \texttt{\_handle\_directory}, \texttt{\_write\_file} and \texttt{\_write\_directory} functions need the \texttt{wpath} promise, while the \texttt{\_handle\_text} function does not need any permissions at all. The Threadbox sandbox for these functions is shown in Listing \ref{lst:Sandboxing wormhole receive functions}.

While being very simple, Threadbox enables modular sandboxing, as seen in this case study, developers can attach sandboxing to specific functions, each having only the needed permissions for it to run. By being sandboxed, these functions can not perform any networking system calls, or launch new processes. Placing these functions under a sandbox enforces good behaviour, and blocks any permissions the function does not \emph{need}. Any other functions in the program are not affected by these sandboxes, since the sandbox is only applied to the annotated~functions.

\code{Sandboxing wormhole receive functions}{Python}{code/wormhole_sandboxed.py}

Deciding which functions are a good candidate for sandboxing can be challenging especially if the code is not well organized and functions perform more than one task. However, splitting the code into different functions is much easier than re-architecting the application to use a multi-process model for instance. As we mentioned earlier, modularity promotes security, and the more modular the code is, the easier it becomes to sandbox it. An argument can be made that our approach influences developers to write better code and organize their code to be more modular by developing functions that are responsible for performing specific tasks.

To evaluate how Threadbox enhances security, we created a simulated vulnerability designed to mimic a backdoor that would be triggered if a transferred file had a specific name. This models a scenario in which a malicious maintainer embeds exploitable code within the application. The goal is to test whether Threadbox’s sandboxing can effectively block such a backdoor from compromising the system.

The backdoor code is shown in Listing~\ref{lst:Backdoor function}. It checks the filename, extracts an encoded command, decodes it, and then attempts to execute it. This function is invoked from within the \texttt{\_handle\_file} function, which is shown in Listing~\ref{lst:Handle file function}.

\code{Backdoor function}{Python}{code/backdoor.py}

\code{Handle file function}{Python}{code/wormhole_handlefile.py}

When transferring regular files, the \texttt{\_handle\_file} function executes successfully and only requires the \texttt{wpath} promise. However, if a file is sent with a specially crafted name that triggers the backdoor, the function will attempt to execute a command and require additional privileges specifically, the \texttt{proc} promise. In this case, Threadbox should prevent the exploit.

To test this, we sent a file with a file name that activates the backdoor and attempts to disclose the contents of \texttt{/etc/passwd}, as shown in Listing~\ref{lst:File name that results in triggering the backdoor}. When this file is received, Threadbox detects that the sandboxed function is attempting to use the \texttt{proc} permission which is not granted, and immediately terminates the process. The kernel logs generated during this event are shown in Listing~\ref{lst:Threadbox logs when the backdoor is triggered}.

\code{File name that results in triggering the backdoor}{Lisp}{code/backdoor_command.txt}

\code{Threadbox logs when the backdoor is triggered}{Lisp}{code/backdoor_logs.txt}

\subsection{Case Study: Sandboxing a Vulnerable PDF Reader}

This case study explores how Threadbox can be used to sandbox a vulnerable Java-based PDF reader. We developed a simple PDF reader using Java Swing and the \texttt{itext} library, a widely-used PDF parser for both Java and C\#. The version of \texttt{itext} used in this case is known to be vulnerable to an XXE (XML External Entity) attack, documented as \texttt{CVE-2017-9096}.

Our implementation features a basic GUI built with Swing, allowing users to open and view the contents of PDF files. When a file is selected, the application spawns a new thread to parse the file and display its textual content in a separate window.

To test whether Threadbox can effectively block exploits delivered through malicious PDFs, we crafted a document that exploits the XXE vulnerability in \texttt{itext}. This document attempts to initiate communication with an attacker-controlled server, simulating a real-world exploit~scenario.

The PDF reader program utilises multiple threads to manage the GUI and parse PDF files. Sandboxing this application using Seccomp presents a number of challenges. First, there are no native Seccomp bindings for Java, meaning developers would need to write C code and integrate it using the Java Native Interface (JNI). Second, the Java Virtual Machine (JVM) issues a large number of system calls during startup and runtime, making it difficult to isolate and determine the specific system calls required by just the PDF parser thread.

When approaching this with Seccomp, the parser thread requires access to 25 different system calls. Identifying these system calls involved a multi-step and error-prone process. First, we added logging code to the Java application and rebuilt it into a JAR file. Then, we configured \texttt{strace} to generate separate output files for each thread, and ran the JAR under \texttt{strace}. Afterwards, we identified the output file corresponding to the parser thread and filtered it using \texttt{awk} to extract the system calls, as shown in Listing~\ref{lst:AWK command to filter the system calls needed by the parser thread}.  All of this effort was required just to determine the necessary system calls.

\code{AWK command to filter the system calls needed by the parser thread}{Bash}{code/awk_script.txt}

Threadbox simplifies this process through its high-level abstraction layer, which avoids dealing with the low-level system call interface altogether. To determine the permissions required by the parser thread, developers can use the \texttt{permissions} function, assign a descriptive name to the sandbox (making it easier to identify in the logs), and enable learning mode. This process revealed that the parser thread only requires the \texttt{rpath} and \texttt{unix} promises. Threadbox allows developers to attach names to their sandbox, making it easier to identify in the logs, and also provides a learning mode, that allows figuring out what promises a component needs. Developers can configure these settings \emph{inside the code}, they do not have to use external tools, or write custom scripts to identify the sandbox policy.

The parser thread can then be sandboxed as shown in Listing~\ref{lst:PDF parser sandbox}, where only the \texttt{rpath} and \texttt{unix} promises are granted, allowing the thread to read the PDF file and display its contents using the GUI. If the reader attempts to use any other permissions, the process is terminated by~Threadbox.

\code{PDF parser sandbox}{Java}{code/pdf_viewer.java}

In this case study, the PDF reader program consists of two threads: the main thread, which creates the GUI, and the parser thread, which parses the PDF and displays it in a new window. Threadbox allows developers to attach different sandboxes to each thread independently. For example, the main thread of the PDF reader program can be sandboxed using \texttt{permissions("unix rpath threading")}. The \texttt{unix} promise is required to create the GUI, \texttt{rpath} is needed to open the file chooser, and \texttt{threading} is necessary to launch new threads (in this case, the parser thread). Because Threadbox enables attaching independent sandboxes to individual threads, the main thread is allowed to use \texttt{threading}, while the parser thread will not have access to this~permission.

\code{PDF viewer main thread}{Java}{code/main_pdf.java}

The PDF reader program in this case study uses a popular library to parse PDF files \texttt{itext}. This library was vulnerable to XXE attacks, where an attacker can craft a malicious PDF file that triggers a network request from the victim's system. This simple vulnerability that is caused by a PDF parser making a networking request when opening a PDF file can be leveraged to cause a code execution. For instance, an internal vulnerable HTTP server that is supposed to be accessed only from within a DMZ can be exploited using an XXE to gain remote code execution by just simply opening a crafted PDF using a vulnerable PDF reader, which will then send a crafted request to the internal vulnerable server exploiting it. Sandboxing and limiting the permissions of PDF readers can help in mitigating such vulnerabilities.

To illustrate how Threadbox protects against these attacks, we used a vulnerable version of the \texttt{itext} library and tested the PDF reader program twice: once with a normal PDF file, and once with a malicious one. The malicious file contains embedded XML forms with an XXE entity that attempts to communicate with an external attacker-controlled server, as shown in Listing~\ref{lst:XXE entity in the malicious PDF file}.

\code{XXE entity in the malicious PDF file}{Lisp}{code/xxe.txt}

When opening the normal file, the PDF reader successfully parses the content and displays it in a new window. However, when opening the malicious file, the PDF reader is terminated by Threadbox, and the incident is reported in the logs, as shown in Listing~\ref{lst:Threadbox logs when opening the malicious PDF file}.

\code{Threadbox logs when opening the malicious PDF file}{Lisp}{code/xeelogs.txt}

The XXE exploit causes the PDF reader to attempt a network request, which requires the \texttt{net} promise. Since this permission is not granted to the parser thread, Threadbox prevents the action and kills the process to stop the attack. This sandboxing model also protects against other known PDF vulnerabilities that attempt to execute system commands, as the parser thread is not granted the \texttt{proc} promise. Even if the PDF reader is running as root, the parser thread is still limited to only the \texttt{rpath} and \texttt{unix}~promises.

\section{Performance Benchmarks}

Threadbox is technically simple and does not perform any complex computations. However, any code running in the kernel must not cause performance issues, as this would halt the whole system. As Threadbox is still in very early stages and is far from optimal, the goal of the performance tests in this section is to examine whether Threadbox in its current state results in significant performance issues or not, and to notice if there are any bottlenecks in its design. An initial evaluation
of Threadbox performance is needed to better understand the overhead it introduces on the sandboxed applications and the wider~system.

We performed multiple performance benchmarks on Threadbox to evaluate its performance in different cases. The tests were carried out on a Linux 6.4.10 system with an x86\_64 2700Mhz 12th Gen Intel CPU.

In the first case study we presented earlier, Threadbox was used to sandbox a Flask web server. When deployed in production environments, such web servers could handle thousands of requests per second in parallel, and if a security mechanism results in slowing down the response time of the server, this could lead to developers abandoning it. We measured the performance overhead Threadbox introduces to a Flask web server by using the Apache HTTP server benchmarking tool \cite{ab_tool}. Table \ref{tab:webserver_benchmarks} shows the benchmarks of the server before and after applying the Threadbox sandbox. The server was tested by issuing 1000 requests in parallel to an endpoint and measuring the response time for each request. The results were repeated again but with the Threadbox sandbox applied to the endpoint as shown in the case studies section.

The results show that Threadbox introduces minimal overhead. The time taken for the whole test to finish was different by only 300 ms, while the median response time per request increased by 2 ms. The throughput of the server decreased by 13\% with a difference of 55 requests per second during the whole test time.
When measuring the response time, we used the median average as it is more robust against outliers. Imagine measuring the execution time of 1000 requests. If the majority of requests take about 10 ms, but a few outliers take 500 ms due to occasional system delays (like a slow CPU interrupt), then the mean average will increase due to the noise spike, while the median will remain close to 10 ms.

\begin{table}
  \centering
  \caption{Response time of a Flask API endpoint before and after the sandbox. The requests per second show the mean average, while the time per request show the median~average.}
    \begin{tabular}{lccc}
          \toprule
          Measurement & Before & After & Difference \\
          \midrule
          Time taken for test (s) & 2.176 & 2.477 & +0.301 (+13.83\%) \\
          Requests per second  & 459.55 & 403.77 & -55.78  (-13.81\%) \\
          Time per request (ms) & 21.760 & 23.767 & +2.00  (+9.19\%) \\
          Transfer rate (kbytes/s) & 2257.81 & 1983.76 & -274.05  (-12.13\%) \\
          \bottomrule \\
      \end{tabular}
    \label{tab:webserver_benchmarks}
\end{table}

Threadbox function-sandbox requires launching new threads for each sandboxed function, in the case of \texttt{magic-wormhole}, we sandboxed five functions as discussed in the previous sections. We ran the different commands of \texttt{magic-wormhole} (sending a text, file, and a directory) 50 times with the same data and measured the time taken for these functions to return results before and after the sandbox. We wrote a shell script that launched \texttt{magic-wormhole} with the needed parameters and timed the results. Table \ref{table:funcsandbox_performance} shows the test results. Using Threadbox to sandbox functions resulted in minimal overhead overall. The sending text functionality was affected the least, as \texttt{\_handle\_text} function does not trigger the \texttt{\_write\_file} or \texttt{\_write\_directory} functions, which are sandboxed. When sending files or directories, two functions will be launched in a Threadbox sandbox, the handler function and the one that writes the results to the disk as shown in Listing \ref{lst:Sandboxing wormhole receive functions}. This results in 2-3\% overhead as shown in the Table. While this case does not show any significant performance overhead, it does not generalize, as more complex applications that require attaching a sandbox to many functions could suffer performance issues.

\begin{table}
  \centering
  \caption{Response time of wormhole receive functions before and after the sandbox. The measurement show the median average.}
    \begin{tabular}{lccc}
          \toprule
          Endpoint & Before (s) & After (s) & Difference\\
          \midrule
          Receive text & 1.420 & 1.444 & +0.024 (+1.69\%)\\
          Receive file & 1.662 & 1.720 & +0.058  (+3.48\%)\\
          Receive folder & 1.771 & 1.813 & +0.042  (+2.37\%)\\
          \bottomrule \\
      \end{tabular}
    \label{table:funcsandbox_performance}
\end{table}

If Threadbox LSM is installed on a system, then each system call has to go through the LSM hooks registered by Threadbox, even if the process issuing these system calls is not sandboxed, and then it is for the LSM to determine whether to act on the triggered system call or not. If the process of checking whether to act on a system call or not is slow, then all programs running this system call, regardless of being sandboxed or not, will be affected.

To measure the effects of Threadbox default checks on system calls, we took micro-benchmarks of various system calls before and after compiling Threadbox into the kernel. Table \ref{tab:micro_bencmarks} shows micro-benchmarks of individual system calls on kernels before and after adding Threadbox LSM. We utilized \texttt{libmicro} (developed by Red Hat) to run each system call 1000 times and then compare the results \cite{libmicro_github}.

\begin{figure}[H]
    \centering
    \includegraphics[scale=.42]{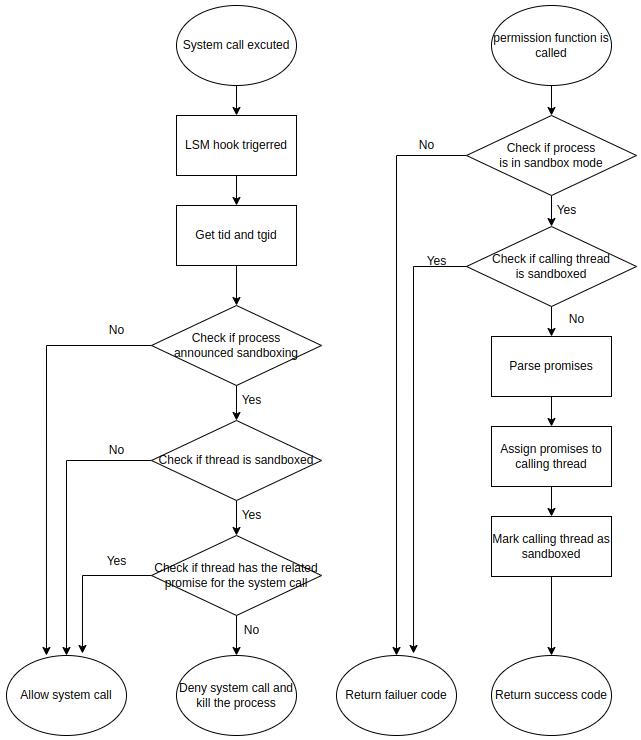}
    \caption{Threadbox LSM flow diagram}
    \label{fig:flow_diagram}
\end{figure}

The Table shows a variety of system calls. The \texttt{getpid} system call is executed by the VDSO and does not reach the kernel, thus it is not affected by Threadbox LSM hooks (only 0.01 microseconds which we consider to be noise). Similarly, the \texttt{lseek} system call is not affected because is not hooked by Threadbox, and it is not checked even if Threadbox is installed. There is a slight performance overhead with system calls that are hooked by Threadbox, such as \texttt{connect}, \texttt{listen}, and \texttt{open}. The \texttt{socket\_i} and \texttt{socket\_u} corresponds to \texttt{inet} and \texttt{unix} sockets. The \texttt{open} system call is tested by opening a \texttt{tmp} file, user-created file, and \texttt{/dev/zero}, which does not read from the hard desk, but rather, feeds the system call with constant null values. Overall, Threadbox does not impose significant performance issues on programs running on a system compiled with it, because the checking process is very simple, and processes that are not sandboxed do not go through the full checking process as shown in the flow diagram (Figure \ref{fig:flow_diagram}).

\begin{table}
  \centering
  \caption{Micro-benchmarks of various system calls before and after compiling Threadbox into the kernel}
    \begin{tabular}{lccc}
      \toprule
      System call & Before (µs) & After (µs) & Difference\\
      \midrule
      \texttt{connect} & 2.69 & 2.92 & +0.23 (+8.4\%)\\
      \texttt{listen} & 0.56 & 0.61 & +0.05 (+10.5\%)\\
      \texttt{accept} & 1.70 & 1.96 & +0.26 (+15.1\%)\\
      \texttt{socket\_i} & 1.80 & 1.96 & +0.16 (+8.4\%)\\
      \texttt{socket\_u} & 1.63 & 1.79 & +0.16 (+9.8\%)\\
      \texttt{socketpair} & 2.61 & 2.88 & +0.27 (+10.2\%)\\
      \texttt{open\_tmp} & 1.32 & 1.44 & +0.12 (+8.8\%)\\
      \texttt{open\_usr} & 1.40 & 1.48 & +0.08 (+6.1\%)\\
      \texttt{open\_zero} & 1.31 & 1.35 & +0.04 (+3.0\%)\\
      \texttt{getpid} & 0.30 & 0.29 & -0.01 (-4.3\%)\\
      \texttt{lseek} & 0.51 & 0.51 & 0.0 (0.0\%)\\
      \bottomrule
    \end{tabular}
  \label{tab:micro_bencmarks}
\end{table}

\section{Limitations}

Many programming languages and frameworks have lightweight threads that are managed by the language runtime, for instance, virtual threads have been introduced to Java recently to implement high-throughput concurrent applications and for launching a large number of concurrent operations without being bound to the limited number of OS threads~\cite{beronic2021analyzing}. Virtual threads are not tied to OS threads and are managed by the JVM, where many virtual threads can be running under the same OS thread. Similar concepts exist in Nodejs and Golang. Threadbox can not be used with virtual threads, since this concept does not exist in kernel space, and we can not identify virtual threads in kernel code.

Threadbox makes it easy to sandbox specific components in an application, however, this comes at a cost. With BPF filters, developers can implement strict and fine-grained sandboxing by inspecting system call arguments and can for instance limit sockets to only operate on specific protocols, additionally, when a promise is granted using Threadbox, it allows many system calls to run, and some of these system calls might not be needed by an application. Threadbox does not have such capabilities and instead, focuses on simplicity. Seccomp inheritance feature provides a layer of security where a thread can not gain more privilege than its parent, however, as discussed earlier, this often results in the need for re-architecture.

Threadbox compromises having full security and allows for attaching independent sandboxes without inheritance. This approach can be abused by attackers if the \texttt{proc} promise is granted as attackers can launch new processes that do not have sandboxing features and carry their attacks there. However, we argue that our model's aim is not to provide full security but rather to reduce the attack surface. When developers grant the \texttt{proc} promise using Threadbox, they allow the thread to launch new \emph{unconfined} processes, however, since Threadbox does not propagate the parent sandbox to its children, developers can sandbox these threads or processes using a sandbox that fits them better than their parent sandbox. A similar limitation exists in the Pledge model since the Pledge sandbox is not inherited by children process (unless \texttt{execpromises} argument is set) and if the \texttt{proc} promise is granted, attackers can launch new processes without sandbox protection. Similarly, granting the \texttt{threading} promise allows launching unconfined threads, but this does not enable running external commands like the shell for instance, and only allows running internal code (functions) in new threads.

Another weakness in Threadbox model is that threads share the same memory space, since all of the threads in a process are contained within the same virtual address, a code in one thread can still access global variables and file descriptors that can be used by other threads. However, we believe that imposing restrictions on what a thread can do with this data provides a layer of security. If a sandbox restricts the thread from using the \texttt{write} system call, then the thread will not be able to use this system call with an opened file descriptor by another thread. If the code is originally thread-unsafe, Threadbox does not solve such a problem, as this is not its goal, but it reduces the attack surface and limits what attackers can do in case of a compromise.

The shared memory space between threads imposes limitations on what a sandbox can do, for instance, Seccomp does not allow inspecting pointer arguments due to Time-of-check-to-time-of-use (TOCTTOU) attacks, if one thread uses a pointer as an argument to a system call, another thread can modify the value of this pointer, and it would not be pointing to the same location at the time of triggering the system call. This limitation does not apply to Threadbox sandbox since we do not allow inspecting arguments in the first place.

One of the key limitations of this work is not directly related to Threadbox itself, but rather to the design of the Linux operating system. Since the kernel space and user space are developed by separate teams, it becomes challenging to determine when to grant permissions in the kernel. For instance, in systems that use \texttt{systemd}, DNS requests require the \texttt{unix} promise to communicate with the resolver service. However, on other types of systems, this permission may not be necessary. In contrast, systems like OpenBSD do not face this issue, as both the user space and kernel space are developed by the same team. This problem not only requires collaboration from operating system maintainers, but also from application developers. For example, we observed that the Flask platform sometimes communicates with the \texttt{uuidd} local service when the server crashes. In these cases, the varied use of \texttt{unix} sockets introduces challenges, as it becomes difficult to determine when to grant such permissions.

To store the details of sandboxed processes and threads, Threadbox uses fixed-size arrays. This creates two limitations. First, the number of sandboxed processes and threads can not exceed the array sizes. Second, the process of looking up if a process or a thread is sandboxed will take O(N) time at most. This time can be reduced further by using a more efficient data structure than arrays, such as a Hash Table, or a better searching algorithm than linear search, such as binary search, leading to better performance overall when checking if a process or thread is sandboxed.

Threadbox is still in early stages of development, and currently, the list of system calls that are checked might not cover all existing architectures and edge cases. Future work includes covering more architectures, placing hooks on additional system calls, and possibly, making the promises context-aware as in the Pledge model in~OpenBSD.

\section{Related work}

The field of sandboxing encompasses a wide range of diverse research. Maass et al. performed an interdisciplinary systematic literature review, offering a formal definition of the term `sandbox' \cite{maass2016systematic}. They also highlighted several research gaps, particularly in the areas of validation techniques and usability. Similarly, Schreuders et al. conducted a literature review, classifying sandboxes into two categories, rule-based and isolation-based \cite{schreuders2013state}. There are also many research that touched on system and application containers~\cite{bhat2022introduction, docker20164}

In contrast to the broader sandboxing research, studies on sandboxing APIs such as Seccomp, Pledge, and Capsicum, are less extensive. Listing \ref{lst:Seccomp and Pledge examples} demonstrates an example of a Seccomp filter that permits only the \texttt{write} system call, along with a Pledge example that allows the \texttt{stdio} promise.

\code{Seccomp and Pledge examples}{C}{code/seccomp_example.c}

Research by Anderson provides a comparison of Seccomp, Pledge, and Capsicum, and examines thier implementation details~\cite{anderson2017comparison}. Other research focuses on utilizing these mechanisms to implement other sandboxing solutions, such as the research by Bijlani et al. that uses Seccomp to build SandFS, a file system sandbox that allows for having a limited view of the file system for the sandboxed process~\cite{capsicum-cap-for-unix}, and research by Jadidi et al. that proposes CapExec, a sandbox for services that uses Capsicum under the hood~\cite{jadidi2019capexec}.  There is also research that focuses on automatically generating policies for Seccomp~\cite{canella2021automating, lopes2020container, xing2022devil}.

Alhindi et al. analyzed the use of sandboxing mechanisms in Debian, Fedora, OpenBSD, and FreeBSD packages by identifying packages that implement Seccomp, Pledge, or Capsicum, and comparing how these mechanisms perform across different operating systems for the same packages. Their findings showed that only around 1\% of packages use these mechanisms directly across the different operating systems, with Pledge having the highest percentage. They also observed that developers sometimes use fine-grained mechanisms like Seccomp to emulate simpler models like Pledge. The research highlighted significant challenges, such as the large amount of code required for Seccomp and the need for software re-architecting to accommodate sandboxing~\cite{alhindi2024sandboxing}.

In another research, Alhindi et al. conducted a usability study on Seccomp with experienced developers and discussed their sandbox implementations and the challenges they encountered. The study highlights many usability issues with Seccomp that stem from the complexity of the system call interface, as some participants ended up with an over-privileged sandbox that allows more system calls than needed. In the design phase of the study, some participants suggested a sandbox that could \emph{wrap} functions, since this approach would create more locality between the code and the sandbox policy. Threadbox builds on this research and addresses many of the usability issues found with Seccomp, such as the complexity of the sandbox policy and the debugging interface.

\section{Conclusion}

In this work, we examine what makes current sandboxing solutions difficult to apply to common applications like web servers, and propose Threadbox, a sandboxing mechanism that is designed to offer \emph{independent} and \emph{modular} sandboxing. We show its technical design and explain the reasoning behind our decisions, showcase how Threadbox can be used to sandbox common applications with ease, illustrate how it can be used to apply a sandbox for Python functions, and touch on its limitations. This work proposes a new approach for sandboxing that addresses many challenges and enables developers to achieve the principle of the least privilege in their code easily.

\bibliographystyle{IEEEtran}
\bibliography{IEEEabrv,bib}

\begin{thebibliography}{10}
\providecommand{\url}[1]{#1}
\csname url@samestyle\endcsname
\providecommand{\newblock}{\relax}
\providecommand{\bibinfo}[2]{#2}
\providecommand{\BIBentrySTDinterwordspacing}{\spaceskip=0pt\relax}
\providecommand{\BIBentryALTinterwordstretchfactor}{4}
\providecommand{\BIBentryALTinterwordspacing}{\spaceskip=\fontdimen2\font plus
\BIBentryALTinterwordstretchfactor\fontdimen3\font minus
  \fontdimen4\font\relax}
\providecommand{\BIBforeignlanguage}[2]{{%
\expandafter\ifx\csname l@#1\endcsname\relax
\typeout{** WARNING: IEEEtran.bst: No hyphenation pattern has been}%
\typeout{** loaded for the language `#1'. Using the pattern for}%
\typeout{** the default language instead.}%
\else
\language=\csname l@#1\endcsname
\fi
#2}}
\providecommand{\BIBdecl}{\relax}
\BIBdecl

\bibitem{alhindi2024sandboxing}
M.~Alhindi and J.~Hallett, ``Sandboxing adoption in open source ecosystems,''
  in \emph{Proceedings of the 12th ACM/IEEE International Workshop on Software
  Engineering for Systems-of-Systems and Software Ecosystems}, 2024, pp.
  13--20.

\bibitem{Maass-thesis}
M.~Maass, ``A theory and tools for applying sandboxes effectively.'' Ph.D.
  dissertation, Carnegie Mellon University, USA, 2016.

\bibitem{Omar_Polo_2021}
\BIBentryALTinterwordspacing
O.~Polo, ``Comparing sandboxing techniques,'' 2021. [Online]. Available:
  \url{https://www.omarpolo.com/post/gmid-sandbox.html}
\BIBentrySTDinterwordspacing

\bibitem{duan2020towards}
R.~Duan, O.~Alrawi, R.~P. Kasturi, R.~Elder, B.~Saltaformaggio, and W.~Lee,
  ``Towards measuring supply chain attacks on package managers for interpreted
  languages,'' \emph{arXiv preprint arXiv:2002.01139}, 2020.

\bibitem{vaidya2019security}
R.~K. Vaidya, L.~De~Carli, D.~Davidson, and V.~Rastogi, ``Security issues in
  language-based software ecosystems,'' \emph{arXiv preprint arXiv:1903.02613},
  2019.

\bibitem{flask-doc}
\BIBentryALTinterwordspacing
{Flask project}, ``Flask documentation,'' 2023. [Online]. Available:
  \url{https://flask.palletsprojects.com/en/3.0.x/api/#flask.request}
\BIBentrySTDinterwordspacing

\bibitem{Abbadini_Facchinetti_Oldani_Rossi_Paraboschi_2023}
M.~Abbadini, D.~Facchinetti, G.~Oldani, M.~Rossi, and S.~Paraboschi,
  ``\BIBforeignlanguage{en}{Cage4deno: A fine-grained sandbox for deno
  subprocesses},'' in \emph{\BIBforeignlanguage{en}{Proceedings of the ACM Asia
  Conference on Computer and Communications Security}}.\hskip 1em plus 0.5em
  minus 0.4em\relax Melbourne VIC Australia: ACM, Jul. 2023, p. 149–162.

\bibitem{capsicum-cap-for-unix}
R.~N.~M. Watson, J.~Anderson, K.~Kennaway, and B.~Laurie,
  ``\BIBforeignlanguage{en}{Capsicum: practical capabilities for unix},''
  p.~17, 2010.

\bibitem{alhindi2025playing}
M.~Alhindi and J.~Hallett, ``Playing in the sandbox: A study on the usability
  of seccomp,'' \emph{Twenty-First Symposium on Usable Privacy and Security},
  2025.

\bibitem{schreuders2012towards}
Z.~C. Schreuders, T.~McGill, and C.~Payne, ``Towards usable
  application-oriented access controls: qualitative results from a usability
  study of {SELinux}, {AppArmor} and {FBAC-LSM},'' \emph{International Journal
  of Information Security and Privacy (IJISP)}, vol.~6, no.~1, pp. 57--76,
  2012.

\bibitem{Tunney_pledge_2022}
\BIBentryALTinterwordspacing
J.~Tunney, ``\BIBforeignlanguage{en}{Porting {OpenBSD} pledge() to {Linux}},''
  2022. [Online]. Available: \url{https://justine.lol/pledge/}
\BIBentrySTDinterwordspacing

\bibitem{serenityos_github}
\BIBentryALTinterwordspacing
A.~Kling, ``Serenity {OS} github page,'' 2024. [Online]. Available:
  \url{https://github.com/SerenityOS/serenity}
\BIBentrySTDinterwordspacing

\bibitem{Jake_2010}
\BIBentryALTinterwordspacing
J.~Edge, ``Restricting the network,'' \emph{LWN.net}, 2010. [Online].
  Available: \url{https://lwn.net/Articles/368730/}
\BIBentrySTDinterwordspacing

\bibitem{corbet2005linux}
J.~Corbet, A.~Rubini, and G.~Kroah-Hartman, \emph{Linux device drivers}.\hskip
  1em plus 0.5em minus 0.4em\relax O'Reilly Media, Inc., 2005.

\bibitem{flask_dashboard}
\BIBentryALTinterwordspacing
{ App Generator}, ``{Datta Able} - open-source {Flask} dashboard,'' 2025.
  [Online]. Available: \url{https://github.com/app-generator/flask-datta-able}
\BIBentrySTDinterwordspacing

\bibitem{pyyaml}
\BIBentryALTinterwordspacing
K.~Simonov, ``Pypi stats for {PyYAML},'' 2025. [Online]. Available:
  \url{https://pypistats.org/packages/pyyaml}
\BIBentrySTDinterwordspacing

\bibitem{pyyaml_question}
\BIBentryALTinterwordspacing
Stackoverflow, ``Parsing a yaml file in python, and accessing the data,'' 2013.
  [Online]. Available:
  \url{https://stackoverflow.com/questions/8127686/parsing-a-yaml-file-in-python-and-accessing-the-data}
\BIBentrySTDinterwordspacing

\bibitem{wormhole_github}
\BIBentryALTinterwordspacing
Magic-wormhole, ``Magic-wormhole {Github} page,'' 2024. [Online]. Available:
  \url{https://github.com/magic-wormhole/magic-wormhole}
\BIBentrySTDinterwordspacing

\bibitem{lins2024critical}
M.~Lins, R.~Mayrhofer, M.~Roland, D.~Hofer, and M.~Schwaighofer, ``On the
  critical path to implant backdoors and the effectiveness of potential
  mitigation techniques: Early learnings from xz,'' \emph{arXiv preprint
  arXiv:2404.08987}, 2024.

\bibitem{ab_tool}
\BIBentryALTinterwordspacing
T.~A.~S. Foundation, ``Apache {HTTP} server benchmarking tool,'' 2024.
  [Online]. Available: \url{https://httpd.apache.org/docs/2.4/programs/ab.html}
\BIBentrySTDinterwordspacing

\bibitem{libmicro_github}
\BIBentryALTinterwordspacing
R.~Performance, ``Libmicro performance tool,'' 2024. [Online]. Available:
  \url{https://github.com/redhat-performance/libMicro}
\BIBentrySTDinterwordspacing

\bibitem{beronic2021analyzing}
D.~Beroni{\'c}, P.~Pufek, B.~Mihaljevi{\'c}, and A.~Radovan, ``On analyzing
  virtual threads--a structured concurrency model for scalable applications on
  the {JVM},'' in \emph{2021 44th International Convention on Information,
  Communication and Electronic Technology (MIPRO)}.\hskip 1em plus 0.5em minus
  0.4em\relax IEEE, 2021, pp. 1684--1689.

\bibitem{maass2016systematic}
M.~Maass, A.~Sales, B.~Chung, and J.~Sunshine, ``A systematic analysis of the
  science of sandboxing,'' \emph{PeerJ Computer Science}, vol.~2, p. e43, 2016.

\bibitem{schreuders2013state}
Z.~C. Schreuders, T.~McGill, and C.~Payne, ``The state of the art of
  application restrictions and sandboxes: A survey of application-oriented
  access controls and their shortfalls,'' \emph{Computers \& Security},
  vol.~32, pp. 219--241, 2013.

\bibitem{bhat2022introduction}
S.~Bhat and S.~Bhat, ``Introduction to containerization,'' \emph{Practical
  Docker with Python: Build, Release, and Distribute Your Python App with
  Docker}, pp. 1--9, 2022.

\bibitem{docker20164}
T.~DOCKER, ``4. from chroot over containers,'' \emph{Future Internet (FI) and
  Innovative Internet Technologies and Mobile Communications (IITM)}, vol.~2,
  2016.

\bibitem{anderson2017comparison}
J.~Anderson, ``A comparison of {UNIX} sandboxing techniques,'' \emph{FreeBSD
  Journal}, 2017.

\bibitem{jadidi2019capexec}
M.~S. Jadidi, M.~Zaborski, B.~Kidney, and J.~Anderson, ``Capexec: Towards
  transparently-sandboxed services (extended version),'' \emph{arXiv preprint
  arXiv:1909.12282}, 2019.

\bibitem{canella2021automating}
C.~Canella, M.~Werner, D.~Gruss, and M.~Schwarz, ``Automating {Seccomp} filter
  generation for {Linux} applications,'' in \emph{Proceedings of the 2021 on
  Cloud Computing Security Workshop}, 2021, pp. 139--151.

\bibitem{lopes2020container}
N.~Lopes, R.~Martins, M.~E. Correia, S.~Serrano, and F.~Nunes, ``Container
  hardening through automated {Seccomp} profiling,'' in \emph{Proceedings of
  the 2020 6th International Workshop on Container Technologies and Container
  Clouds}, 2020, pp. 31--36.

\bibitem{xing2022devil}
Y.~Xing, J.~Cao, K.~Sun, F.~Yan, and S.~Wan, ``The devil is in the detail:
  Generating system call whitelist for {Linux Seccomp},'' \emph{Future
  Generation Computer Systems}, vol. 135, pp. 105--113, 2022.

\end{thebibliography}

\appendices

\section{LSM flow diagram}

\end{document}